**Exoplanet Diversity in the Era of Space-based Direct Imaging Missions.**
A white paper submitted in response to the NAS call on
**Exoplanet Science Strategy**

Ravi Kopparapu[1,2,3], Eric Hebrard[4], Rus Belikov[5], Natalie M. Batalha[5], Gijs D. Mulders[6], Chris Stark[7], Dillon Teal[1], Shawn Domagal-Goldman[1,3], Dawn Gelino[8], Avi Mandell[1], Aki Roberge[1], Stephen Rinehart[1], Stephen R. Kane[9], Yasuhiro Hasegawa[10], Wade Henning[1,2], Brian Hicks[1,2], Vardan Adibekyan[11], Edward W. Schwieterman[3,9], Erika Kohler[1], Johanna Teske[12], Natalie Hinkel[13], Conor Nixon[1], Kevin France[14], William Danchi[1], Jacob Haqq-Misra[15], Eric T. Wolf[14], Scott D. Guzewich[1], Benjamin Charnay[16], Giada Arney[1], Hilairy E. Hartnett[17], Eric D. Lopez[1], Dante Minniti[18], Joe Renaud[19], Vladimir Airapetian[1,20], Chuanfei Dong[21], Anthony D. Del Genio[22], Melissa Trainer[1], Gioia Rau[1,23], Adam Jensen[24], Michael Way[22], Carey M. Lisse[25], Wladimir Lyra[10,26], Franck Marchis[27], Daniel Jontof-Hutter[28], Patrick Young[29], Ray Pierrehumbert[30], Chester E. Harman[22,33], Jonathan Fortney[31], Bill Moore[32], Steven Beckwith[34], Everett Shock[17], Steve Desch[17], Kathleen E. Mandt[25], Noam Izenberg[25], Eric B. Ford[35], Shannon Curry[34], Caleb Scharf[33], Ariel Anbar[17]

**Contact:** Ravi kumar Kopparapu,
  NASA GSFC/University of Maryland College Park
   ravikumar.kopparapu@nasa.gov  225-678-0058

1 NASA GSFC
2 University of Maryland, College Park
3 Virtual Planetary Laboratory
4 University of Exeter, UK
5 NASA Ames
6 University of Arizona, Tucson
7 Space Telescope Science Institute
8 NASA Exoplanet Science Institute
9 University of California, Riverside
10 Jet Propulsion Laboratory, Caltech
11 Institute of Astrophysics and Space Sciences, Portugal
12 Carnegie DTM & Observatories
13 Vanderbilt University
14 University of Colorado, Boulder
15 Blue Marble Space Institute of Science
16 Paris-Meudon Observatory
17 Arizona State University
18 Universidad Andrés Bello, Chile
19 George Mason University
20 American University, Washington D.C
21 Department of Astrophysical Sciences, Princeton University
22 NASA GISS
23 Catholic University of America
24 University of Nebraska at Kearney
25 Johns Hopkins University Applied Physics Laboratory

26 California State University, Northridge
27 SETI Institute
28 Department of Physics, University of Pacific
29 School of Earth and Space Exploration, Arizona State University
30 University of Oxford
31 University of California, Santa Cruz
32 Hampton University
33 Columbia University
34 University of California, Berkeley
35 Pennsylvania State University

**Abstract.** This whitepaper discusses the diversity of exoplanets that could be detected by future observations, so that comparative exoplanetology can be performed in the upcoming era of large space-based flagship missions. The primary focus will be on characterizing Earth-like worlds around Sun-like stars. However, we will also be able to characterize companion planets in the system simultaneously. This will not only provide a contextual picture with regards to our Solar system, but also presents a unique opportunity to observe size dependent planetary atmospheres at different orbital distances. We propose a preliminary scheme based on chemical behavior of gases and condensates in a planet's atmosphere that classifies them with respect to planetary radius and incident stellar flux.

**Areas of Progress Since the New Worlds New Horizons Decadal Survey.**
Within the past two decades, the number of confirmed exoplanets has increased a thousandfold, with thousands more waiting to be confirmed (Mullally et al. 2017). In the past few years alone, several habitable zone (HZ) planets around the nearest M-dwarfs were discovered (Anglada-Escudé et al. 2016; Gillon et al. 2017, Dittman et al. 2017). While the efforts to identify habitable exoplanets are increasingly receiving attention, a wide variety of planetary objects have also been discovered which have no analog in our solar system, such as super-Earths and hot Jupiters. Future large space-based missions will have the capability to detect and characterize a multitude of these planets, along with Earth-like worlds. While there is an intense focus on observing biosignature features on exo-Earths, little attention is given to characterization and expected number of other classes of planets.

**Exoplanet science areas where significant progress will likely be made with current and upcoming observational facilities**
With the upcoming launches of *TESS* and *PLATO*, and recently launched *GAIA* telescope, the next decade will see a vast increase in the number and diversity of exoplanet discoveries. *JWST* will help characterize atmospheres of some of these planets and *WFIRST* will provide performance of starlight suppression technologies needed for future large-aperture space-based direct imaging missions. At the same time, ground-based facilities like ELT and GMT (Rodler & López-Morales 2014) will take the lead in characterizing atmospheres of HZ planets around M-dwarfs. Furthermore, through the study of "model exoplanets" within our Solar system, either by missions or remote observations, a more coherent effort for comparative planetology can be established.

**Exoplanet science areas and key questions that will likely remain after these current and planned missions are completed**
In future searches for exo-Earth candidates around nearby Sun-like stars, we will be able to detect several bright planets (Beckwith 2008). According to Stark et al. (2015), for an 8-meter telescope observing 500 stars, the number of exo-Earth candidates detected is ∼ 20 (see Fig. 1 in Stark et al. 2015), although this is strongly dependent on the value of $\eta_{Earth}$, the fraction of stars that have at least one terrestrial mass/size planet in the habitable zone. If we assume that, on average, every star has a planet of some size (Cassan et al. 2012; Suzuki et al. 2016), then there are ∼ 500 exoplanets of all sizes that can be potentially observed. Not considering the ∼20 exo-Earth candidates, most exoplanets will fall into the "non-Earth" classification, and thus far there is no universally accepted classification system for distinguishing among them. This provides a motivation to devise a scheme based on planetary size and corresponding comparative atmospheric characteristics in order to distinguish features between different classes of non-Earth planets.

*Comparative Exoplanetology*
Comparative exoplanetology, analogous to comparative planetology within our solar system, has the capacity to illuminate trends in planet insolation, density, and mass with atmospheric composition, structure, temperature, and other attributes including the makeup, altitudes, and density of atmospheric condensates.

Multiple observation techniques will provide the best information on exoplanet atmospheres. Optical color-color photometry will provide zeroth order rocky-watery-gas giant discrimination, as it has done for the 1 surface water world, 1 rocky worlds, 2 gas giants, and 2 ice giants of our system (Crow et al 2011, Krissansen-Totton et al. 2016). Transit observations are sensitive to trace gases because of the inherently long transit path lengths, but can only sense the uppermost reaches of a cloudy/hazy atmosphere. Phase curves, sensing directly emitted thermal and reflected light, are thus less vulnerable to truncation by clouds and hazes and may allow access to molecules deeper down.

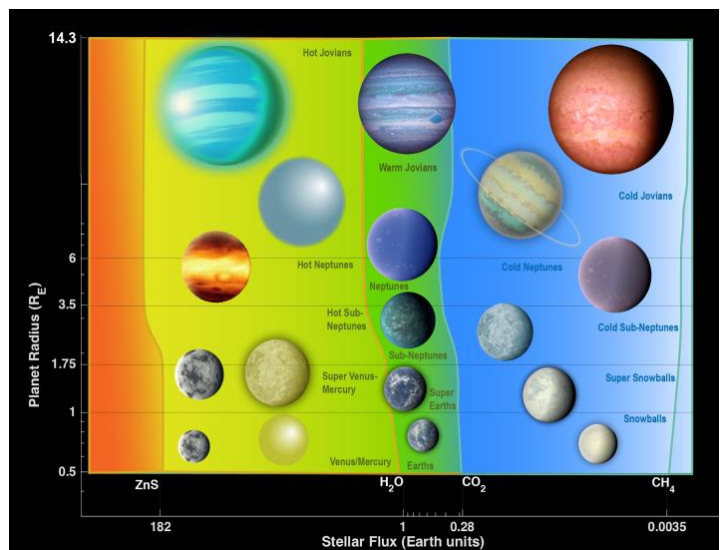

Fig. 1: The boundaries of the boxes represent where different chemical species are condensing in the atmosphere of a planet at a stellar flux, according to equilibrium chemistry calculations (Kopparapu et al. 2018)

Direct imaging in reflected light as a function of orbital phase at UV-optical-NIR wavelengths will provide access to broadband UV absorption features, other visible and NIR absorbers, and evidence of surface composition and liquid surface water. In the end, the most complete understanding of an exoplanet will combine multiple observation techniques across the widest wavelength range.

There is already a need for a coordinated cross-disciplinary effort to interpret spectroscopic observations (Charnay et al. 2015, Hammond & Pierrehumbert 2017). Currently, comparative exoplanetology studies are limited by both sensitivity and spectral resolution to some bulk properties of a collection of mostly hot Jupiter-type planets (e.g., Barstow et al. 2016). JWST will have the capacity to vastly expand exoplanetology to various other planet categories here, including hot, warm, and cold Neptunes and sub-Neptunes and potentially a limited number of terrestrial planets, including super-Earths, Earths, and Venus analogs (Cowan et al. 2015; Greene et al. 2016). JWST will also substantially increase the sensitivity and spectral information obtained for hot, warm, and cold Jovians. Combined with statistical information from WFIRST, this comprehensive dataset will allow unpresented insight into compositional trends and will impose meaningful new constraints planet formation models (e.g., Espinoza et al. 2017; Venturini et al. 2016). The planned European ARIEL mission will allow us to extract the chemical fingerprints of gases and condensates in the planets' atmospheres, including the elemental composition. Near- and mid-IR space interferometry missions to directly image exoplanets around nearby solar type stars will provide information about molecular features from bands of molecules such as carbon dioxide, water vapor, nitrous oxide, methane, hydroxyl and nitric oxide (Airapetian et al. 2017). Future large-aperture, direct-imaging missions such as LUVOIR and HabEx will extend comparative exoplanet studies to primarily outgassed, high-molecular weight atmospheres, most directly comparable to the inner terrestrial planets of our solar system. Here especially, interdisciplinary collaborations will be required for interpretation of exoplanet data.

**Identifying observational, technological, theoretical, and computational needs for making progress in further understanding exoplanets and exoplanetary systems.**
Classifications from this scheme can be used to estimate exoplanet yields, the number of specific classes of exoplanets detected and spectroscopically characterized by a direct imaging mission. The purpose here is not to test if a single, specific exoplanet lies within the boundaries defined by this scheme. Rather, it is useful in understanding the diversity of exoplanet populations, that could be amenable to characterization by direct imaging missions.

With the exception of Venus-type exoplanets (Kane et al. 2012, Schaefer et al. 2016), there has not been a comprehensive effort to classify planets beyond the HZ (but see Forget & Leconte 2013, Pierrehumbert & Ding 2016). Classifying planets of different sizes based on the transition/condensation of different atmospheric species (Sudarsky et al. 2003; Burrows 2005) at different stellar fluxes provides a physical motivation in estimating exoplanet mission yields, separate from exo-Earth candidate yields (Fig. 1). Additionally, the interior structure and composition of the planet affects the atmospheric temperature profile of the planet (Wordsworth et al. 2018). This, in turn, will impact the condensation of minerals, and therefore structural boundaries and composition of the planet (Hinkel & Unterborn, 2018). The radius estimates for Fig. 1 are obtained from Zahnle & Catling (2017), Fulton et al. (2017) and Chen & Kipping (2017).

In some circumstances a planet's internal heat flux will significantly modify its surface temperature relative to the stellar flux. For very close-in worlds with modest-to-high eccentricity maintained by nearby (resonant or non-resonant) perturbers, tidal heating may contribute millions of times the heat output of the modern Earth, but still only 1-5 K in surface temperature variation (Henning et al. 2009). Such issues become more complex for tidally-locked dayside/nightside thermal dichotomies. Conversely, for some icy exomoons, where insolation is weak, internal heat flux variations matter more, but can still fit the scheme in Figure 1 using a combined internal-solar heat flux axis.

The histograms in Fig. 2 visualize the total scientific impact of the habitable planet candidate survey, along with the several other classes of exoplanets, based on 4-m and 16-m diameter telescopes. The y-axis is the expected total numbers of exoplanets observed (yields; also given by the numbers above the bars). At the large architectures (16m telescopes), one can truly see the diversity in exoplanet yields and further characterize different classes of planets. We note that in general, larger apertures are less sensitive to changes in other parameters (such as contrast ratio) than smaller apertures.

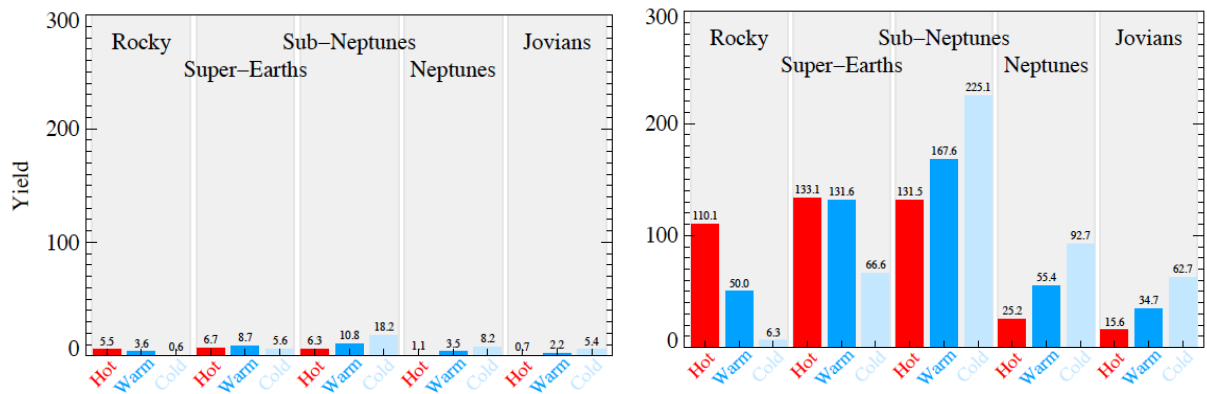

Fig. 2: Expected number of exoplanets observed for hot (red), warm (blue) and cold (ice-blue) incident stellar fluxes shown in Fig. 1. Calculations assume a coronagraph paired with a telescope of diameter 4-m (left) and 16-m (right).

With a 4m class mission, observations designed to maximize the yield of potential Earths will also yield the detection and characterization of all of the planet types discussed here, with the exception of close-in planets. These planets are not observed by a 4m-class mission because of the tight inner working angle.

The 16m class telescopes will bring the ability to not only characterize planets, but also to test the occurrence of different features within each planet type. It would observe dozens of each planet type, providing larger sample sizes which enables the study each planet type as a population. Furthermore, large direct detection missions with UV-to-NIR spectral coverage provides contemporaneous characterization of the host star's high energy irradiance that regulates the atmospheric composition and stability on all types of planets (Harman et al. 2015; Koskinen et al. 2013).

**Identify likely fruitful cross-disciplinary topics and initiatives**
The identification and classification of exoplanet diversity needs expertise in theory and experiments in the planetary science, astrophysics, chemistry, and stellar/heliophysics communities, as well as computational methods and statistical methodologies. Understanding the atmospheric chemical composition and condensation based on the global temperature profile of an exoplanet requires a coordinated cross-disciplinary effort. The vast number of exoplanets that will be available for atmospheric characterization in the near future provides us with a golden opportunity to perform comparative exo-planetology. To that end, a continued support for an agency-wide effort to foster communication and collaborative venue for cross-disciplinary scientists is needed.